\documentclass[aps,prc,reprint,groupedaddress]{revtex4-1}
\usepackage{graphicx} % Include figure files
\usepackage{dcolumn} % Align table columns on decimal point
\usepackage{bm}% bold math
\usepackage{hyperref}% add hypertext capabilities

\begin{document}

\title{Multinucleon transfer dynamics in heavy-ion collisions near Coulomb barrier energies}

\author{Fei Niu$^{1,2}$, Peng-Hui Chen$^{2,3}$, Ya-Fei Guo$^{2}$, Chun-Wang Ma$^{1}$  and Zhao-Qing Feng$^{2}$ }
\email{Corresponding author: fengzhq@impcas.ac.cn}

\affiliation{$^{1}$Institute of Particle and Nuclear Physics, Henan Normal University, Xinxiang 453007, People's Republic of China      \\
$^{2}$Institute of Modern Physics, Chinese Academy of Sciences, Lanzhou 730000, People's Republic of China     \\
$^{3}$University of Chinese Academy of Sciences, Beijing 100190, People's Republic of China}

\date{\today}
\begin{abstract}
The multinucleon transfer reactions near barrier energies has been investigated with a multistep model based on the dinuclear system (DNS) concept, in which the capture of two colliding nuclei, the transfer dynamics and the de-excitation process of primary fragments are described by the analytical formula, the diffusion theory and the statistical model, respectively. The nucleon transfer takes place after forming the DNS and is coupled to the dissipation of relative motion energy and angular momentum by solving a set of microscopically derived master equations within the potential energy surface. Specific reactions of $^{40,48}$Ca+$^{124}$Sn, $^{40}$Ca ($^{40}$Ar, $^{58}$Ni)+$^{232}$Th, $^{40}$Ca ($^{58}$Ni)+$^{238}$U and $^{40,48}$Ca ($^{58}$Ni) +$^{248}$Cm near barrier energies are investigated. It is found that the fragments are produced by the multinucleon transfer reactions with the maximal yields along the $\beta$-stability line. The isospin relaxation is particularly significant in the process of fragment formation. The incident energy dependence of heavy target-like fragments in the reaction of $^{58}$Ni+$^{248}$Cm is analyzed thoroughly.

\begin{description}
\item[PACS number(s)]
25.70.Hi, 24.10.Pa, 24.60.Gv
\end{description}
\end{abstract}

\maketitle

\section{Introduction}

The new isotope synthesis is one of hot topical issues in nuclear physics. It was obtained much progress over past decades and a total of 3224 nuclides has been discovered until the end of 2016 \cite{Th16}. There are several attempts for producing the new isotopes in the different mass domain, i.e., projectile fragmentation reactions, fission of actinide nuclei, fusion-evaporation reactions etc. However, a large area of neutron-rich nuclides with N$>$100  in nuclear chart is blank up to now. The reaction mechanism is particularly important for creating the neutron-rich heavy nuclei in laboratories. The superheavy nuclei (SHN) synthesized via the fusion-evaporation reactions are neutron-deficient and far below the neutron magic number N=184 \cite{Ho15,Og15}. On the other hand, the properties of neutron-rich heavy nuclei are particularly important in understanding the nucleosynthesis during the r-process and the evolutions of shell closure and deformation beyond the neutron number of N=126. The multi-nucleon transfer (MNT) reactions and quasifission (QF) processes have been attempted for producing the heavy neutron-rich nuclei \cite{Ad05,Za07,Za08,Fe09a}. More relaxation time undergoes in the MNT reactions in comparison to the QF process. The shell effects are crucially significant in the low-energy heavy-ion collisions. Several models have been developed for describing the transfer reactions, such as the GRAZING model \cite{Wi94}, dinuclear system (DNS) model \cite{Fe09b,Ad10}, dynamical model based on the Langevin equations \cite{Za15} etc. On the other hand, the microscopic approaches are proposed, i.e., the time dependent Hartree-Fock (TDHF) approach \cite{Go09,Ka13} and improved quantum molecular dynamics (ImQMD) model \cite{Ti08}. The MNT reactions have been investigated in experiments. The damped collisions of two actinide nuclei were investigated at Gesellschaft f\"{u}r Schwerionenforschung (GSI) \cite{Hu77,Sc78,Kr13}. Recently, the attempts to produce the neutron-rich nuclei around N=126 were performed in the reactions of $^{136}$Xe+$^{208}$Pb in Dubna \cite{Ko12} and in Argonne National Laboratory \cite{Ba15}. The MNT reactions of $^{136}$Xe+$^{198}$Pt were also investigated at GANIL \cite{Wa15}. It was found that the MNT reactions have advantages in comparison to the projectile fragmentations \cite{Ku14}.

In this work, the multinucleon transfer mechanism near Coulomb barrier energies is investigated with the dinuclear system (DNS) model. In Sec. II we give a brief description of the DNS model. The isotopic production in MNT is discussed in Sec. III. Summary and perspective on the proton-rich and neutron-rich isotopes are presented in Sec. IV.

\section{Brief description of the DNS model}

The DNS model has been used to describe the formation mechanism of superheavy nuclei in the cold fusion and the $^{48}$Ca induced reactions \cite{Fe06}. The production cross sections of SHN are consistent with the available data and the blank isotopes between the cold and hot fusion reactions and the new elements of Z=119 and 120 are predicted with the proposed projectile-target combinations, incident energies and evaporation channels \cite{Fe10,Fe09}. The multinucleon transfer reactions are to be investigated within the DNS model. The dynamical evolution of colliding system sequentially proceeds the capture process by overcoming the Coulomb barrier, dissipation of several degrees of freedom (relative energy, angular momentum, mass and charge asymmetry etc) within the potential energy surface and the de-excitation of primary fragments. We use the analytical formula, diffusion theory and statistical model for the three stages, respectively. Therefore, the cross sections of the fragments produced in the MNT reactions are evaluated by
\begin{eqnarray}
\sigma_{tr}(Z_{1},N_{1},E_{c.m.})=&& \sum_{J=0}^{J_{\max}}\sigma_{cap}(E_{c.m.},J) \int  f(B)   \nonumber \\
&& \times  P(Z_{1},N_{1},E_{1},J_{1},B)   \nonumber \\
&& \times  W_{sur}(E_{1},J_{1},s) dB,
\end{eqnarray}
where the $E_{1}$ is the excitation energy of the fragment (Z$_{1}$,N$_{1}$). The maximal angular momentum is taken to be the grazing collision of two nuclei. The survival probability $W_{sur}$ of each fragment is evaluated with a statistical approach based on the Weisskopf evaporation theory \cite{Ch16}, in which the excited primary fragments are cooled by evaporating $\gamma$-rays, light particles (neutrons, protons, $\alpha$ etc) in competition with binary fission. The structure effects (shell correction, odd-even effect, Q-value etc) could be particularly significant in the formation of the primary fragments and in the decay process. The capture cross section is calculated within the Hill-Wheeler formula and the barrier distribution approach \cite{Ch17}.

The distribution probability is obtained by solving a set of master equations numerically in the potential energy surface of the DNS. The time evolution of the distribution probability $P(Z_{1},N_{1},E_{1}, t)$for fragment 1 with proton number $Z_{1}$ and neutron number $N_{1}$ and with excitation energy $E_{1}$ is described by the following master equations:
\begin{eqnarray}
&&\frac{d P(Z_{1},N_{1},E_{1}, t)}{dt} =                  \nonumber \\
&& \sum_{Z^{'}_{1}}W_{Z_{1},N_{1};Z^{'}_{1},N_{1}}(t)[d_{Z_{1},N_{1}}P(Z^{'}_{1},N_{1},E^{'}_{1},t)     \nonumber  \\
&& -d_{Z^{'}_{1},N_{1}}P(Z_{1},N_{1},E_{1},t)]        \nonumber \\
&& +\sum_{N^{'}_{1}}W_{Z_{1},N_{1};Z_{1},N^{'}_{1}}(t)[d_{Z_{1},N_{1}}P(Z_{1},N^{'}_{1},E^{'}_{1},t)  \nonumber  \\
&& -d_{Z_{1},N^{'}_{1}}P(Z_{1},N_{1},E_{1}, t)].
\end{eqnarray}
Here the $W_{Z_{1},N_{1};Z^{'}_{1},N_{1}}$($W_{Z_{1},N_{1};Z_{1},N^{'}_{1}}$) is the mean transition probability from the channel($Z_{1},N_{1},E_{1}$) to ($Z^{'}_{1},N_{1},E^{'}_{1}$), [or ($Z_{1},N_{1},E_{1}$) to ($Z_{1},N^{'}_{1},E^{'}_{1}$)], and $d_{Z_{1},Z_{1}}$ denotes the microscopic dimension corresponding to the macroscopic state ($Z_{1},N_{1},E_{1}$).The sum is taken over all possible proton and neutron numbers that fragment $Z^{'}_{1}$,$N^{'}_{1}$ may take, but only one nucleon transfer is considered in the model with the relations $Z^{'}_{1}=Z_{1}\pm1$ and $N^{'}_{1}=N_{1}\pm1$. The motion of nucleons in the interacting potential is governed by the single-particle Hamiltonian \cite{Fe06}.

In the relaxation process of the relative motion, the DNS will be excited by the dissipation of the relative kinetic energy. The local excitation energy is determined by the dissipation energy from the relative motion and the potential energy surface of the DNS as
\begin{eqnarray}
\varepsilon^{\ast}(t)=E^{diss}(t)-\left(U(\{\alpha\})-U(\{\alpha_{EN}\})\right).
\end{eqnarray}
The entrance channel quantities $\{\alpha_{EN}\}$ include the proton and neutron numbers, angular momentum, quadrupole deformation parameters and orientation angles being $Z_{P}$, $N_{P}$, $Z_{T}$, $N_{T}$, $J$, $R$, $\beta_{P}$, $\beta_{T}$, $\theta_{P}$, $\theta_{T}$ for projectile and target nuclei. The excitation energy $E_{1}$ for fragment (Z$_{1}$,N$_{1}$) is evaluated by $E_{1}=\varepsilon^{\ast}(t=\tau_{int})A_{1}/A$. The interaction time $\tau_{int}$ is obtained from the deflection function method \cite{Li83}.

The potential energy surface (PES) of the DNS is given by
\begin{eqnarray}
U(\{\alpha\})=&& B(Z_{1},N_{1})+B(Z_{2},N_{2})-\left[B(Z,N)+V^{CN}_{rot}(J)\right]   \nonumber \\
&& +V(\{\alpha\}).
\end{eqnarray}
The DNS fragments satisfy the relation of $ Z_{1}+Z_{2}=Z $ and  $ N_{1}+N_{2}=N$ with the $Z$ and $N$ being the proton and neutron numbers of composite system, respectively. The symbol ${\alpha}$ denotes the sign of the quantities $Z_{1}$, $N_{1}$, $Z_{2}$, $N_{2}$, $J$, $R$, $\beta_{1}$, $\beta_{2}$, $\theta_{1}$, $\theta_{2}$. The $B(Z_{i},N_{i}) (i=1,2)$ and $B(Z,N)$ are the negative binding energies of the fragment $(Z_{i},N_{i})$ and the compound nucleus $(Z,N)$, respectively. The $V^{CN}_{rot}$ is the rotation energy of the compound nucleus. The $\beta_{i}$ represent the quadrupole deformations of the two fragments. The $\theta_{i}$ denote the angles between the collision orientations and the symmetry axes of the deformed nuclei. The interaction potential between fragments $(Z_{1},N_{1})$ and $(Z_{2},N_{2})$ includes the nuclear, Coulomb, and centrifugal parts \cite{Fe09}. The radial distance $R$ between the centers of the two fragments is chosen to be the value that gives the minimum of the interaction potential, in which the DNS is considered to be formed. Shown in Fig. 1 is the PES in the reaction $^{40}$Ca+$^{248}$Cm. The incident points of projectile and target nuclides are indicated. The nucleon transfer is coupled to the dissipation of relative motion energy and angular momentum starting from the entrance channel. The larger shell correction energy leads to the pocket appearance in the PES. The positive local excitation energy enables the nucleon transition in the valence space. The transition probability is determined by the PES. Therefore, the structure effects of the fragment production in the MNT reactions are embodied in the PES.

%%%%%%%%%%%%%%%%%%%%%%%%%%%%%%%%%%%%% figure 1 %%%%%%%%%%%%%%%%%%%%%
\begin{figure}
\includegraphics[width=8 cm]{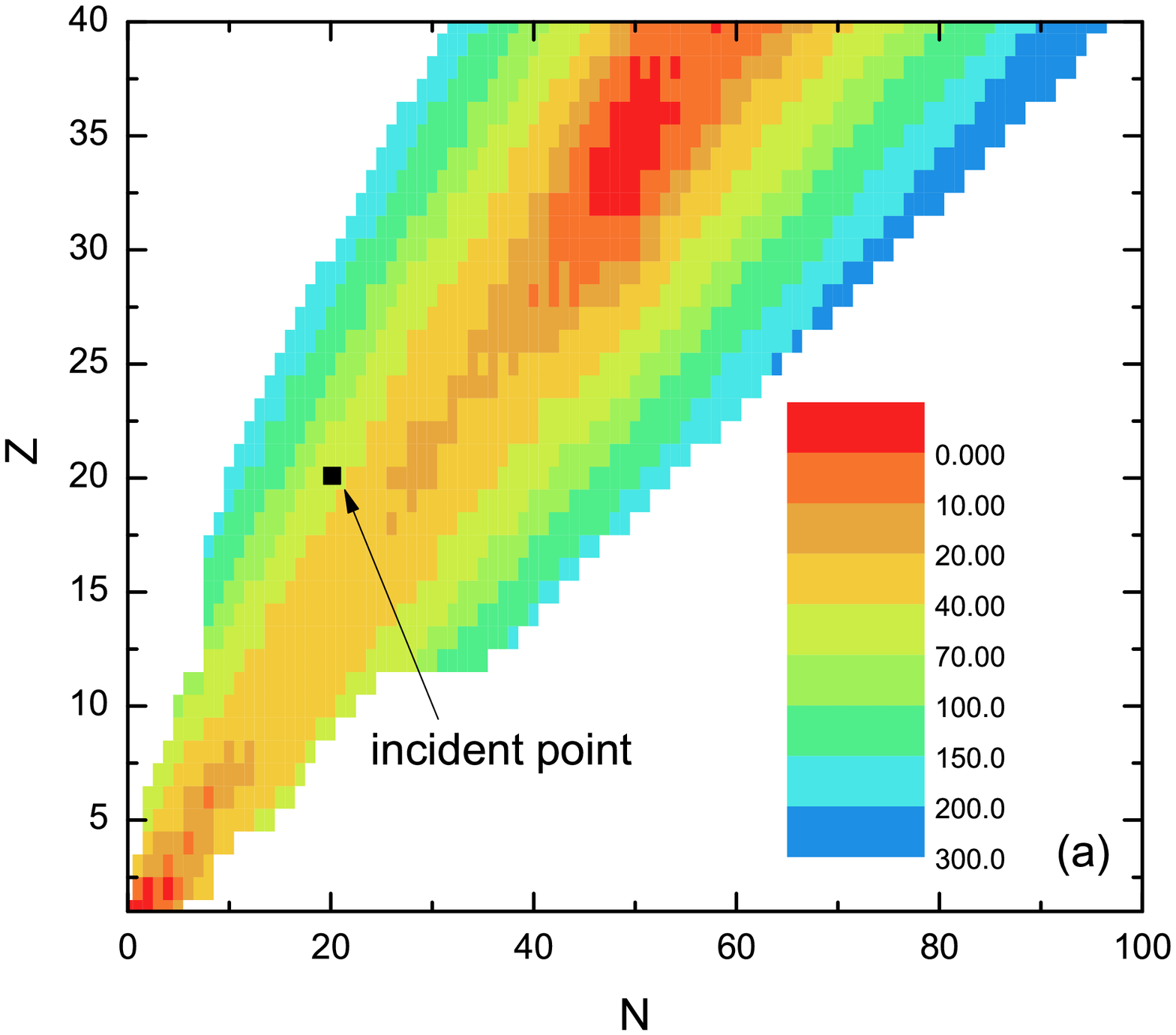}
\includegraphics[width=8 cm]{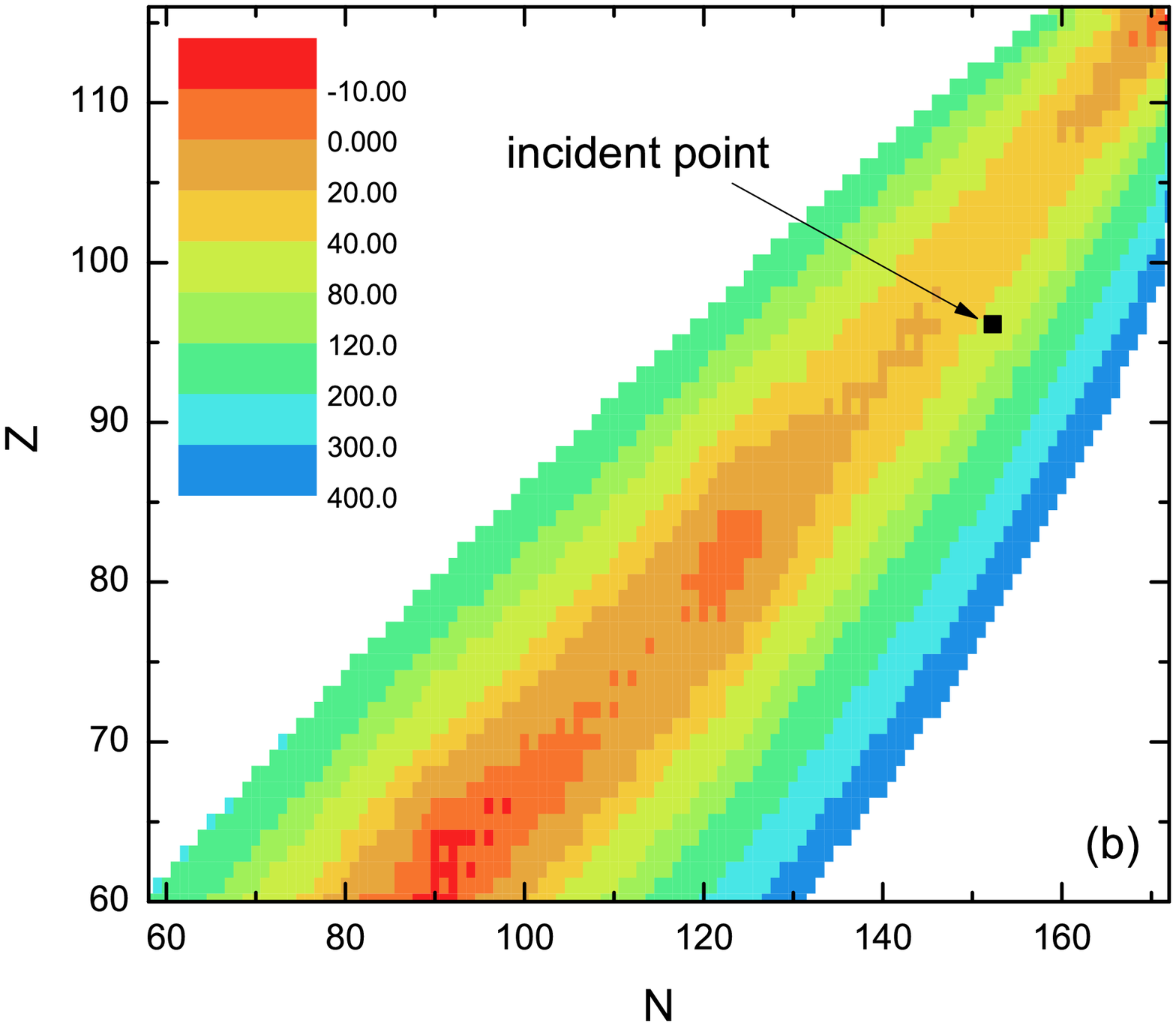}
\caption{(Color online) The potential energy surface in the reaction $^{40}$Ca+$^{248}$Cm as functions of proton and neutron numbers for the projectile-like fragments (upper panel) and target-like fragments (down panel), respectively.}
\end{figure}
%%%%%%%%%%%%%%%%%%%%%%%%%%%%%%%%%%%%%%%%%%%%%%%%%%%%%%%%%%%%%%%

\section{Results and discussion}

In the DNS model, the nuclear structure effects are implemented in the PES, such as the shell effect, odd-even phenomena etc. The dissipation of the collective degrees of freedom is coupled to the nucleon transfer at the touching configuration of two fragments. There is no neck formation in the process of two nuclei approaching. The evolutions of mass and charge distributions of fragments produced in the multinucleon transfer reactions of $^{48}$Ca+$^{124}$Sn are shown in Fig. 2 at the barrier energy ($V_{C}$=111.4 MeV). The reaction system undergoes from the individual nuclei to the broad distributions of projectile-like and target-like fragments with increasing the relaxation time. The new isotopes might be created during the transfer process. The yields are enhanced around the shell closure and the odd-even effect is pronounced.

%%%%%%%%%%%%%%%%%%%%%%%%%%%%%%%%%%%%% figure 2 %%%%%%%%%%%%%%%%%%%%
\begin{figure*}
\includegraphics[width=16 cm]{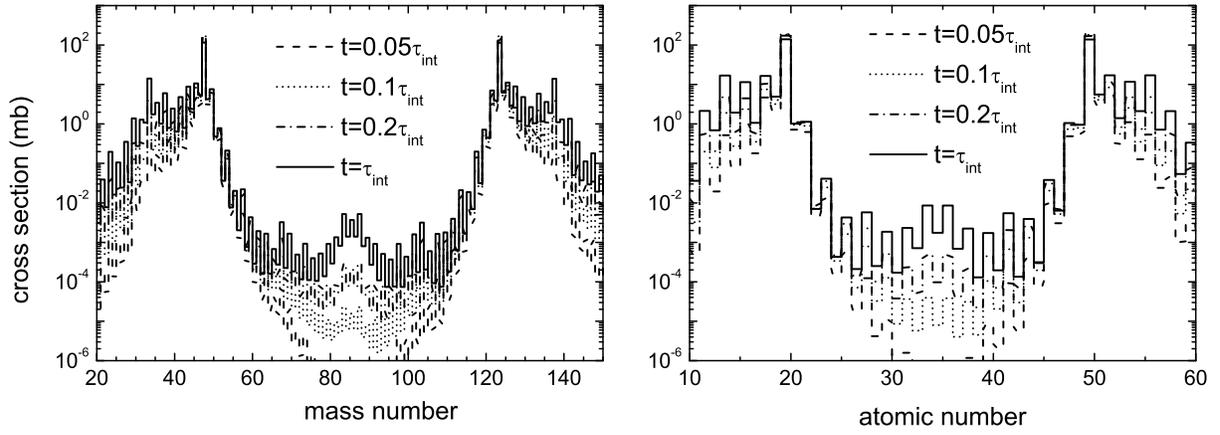}
\caption{\label{fig:wide} Mass and charge distributions of fragments produced in the multinucleon transfer reactions of $^{48}$Ca+$^{124}$Sn at the c.m. energy of 125.4 MeV with the different time step.}
\end{figure*}
%%%%%%%%%%%%%%%%%%%%%%%%%%%%%%%%%%%%%%%%%%%%%%%%%%%%%%%%%%%%%%

As a test of the approach, the projectile-like fragments (PLFs) in the multinucleon transfer reaction $^{40}$Ca+$^{124}$Sn at the energy of 128.5 MeV are calculated and compared with the available data \cite{Co96} as shown in Fig. 3. The isotopes of calcium and potassium are underestimated because of the loss of quasielastic scattering in the model. The yield peaks are produced with neutron-rich nuclei because of the isospin equilibrium, such as $^{37}$Cl, $^{36}$S, $^{33}$P and $^{30}$Si. The reactions with neutron-rich nuclide $^{48}$Ca at the energy of 125.4 MeV ($V_{C}$=111.4 MeV) are investigated as shown in Fig. 4. The experimental data \cite{Co97} are nicely reproduced. Comparison of the target-like fragments (TLFs) in the two reaction systems are shown in Fig. 5. The $^{40}$Ca induced reactions are available for the neutron-deficient nuclide production. The projectile proton stripping enables the formation of heavy fragments. The smaller neutron separation energies of neutron-rich nuclides produced in the multinucleon transfer reactions decreases the survival of the neutron-rich fragments. The neutron shell closure is available for stabilizing the fragments, i.e., possibly producing $^{136}$Xe in the multinucleon transfer reaction of $^{48}$Ca+$^{124}$Sn. Further experiments for checking the shell effect in the transfer reactions are expected.

%%%%%%%%%%%%%%%%%%%%%%%%%%%%%%%%%%%% figure 3 %%%%%%%%%%%%%%%%%%%%%
\begin{figure*}
\includegraphics[width=16 cm]{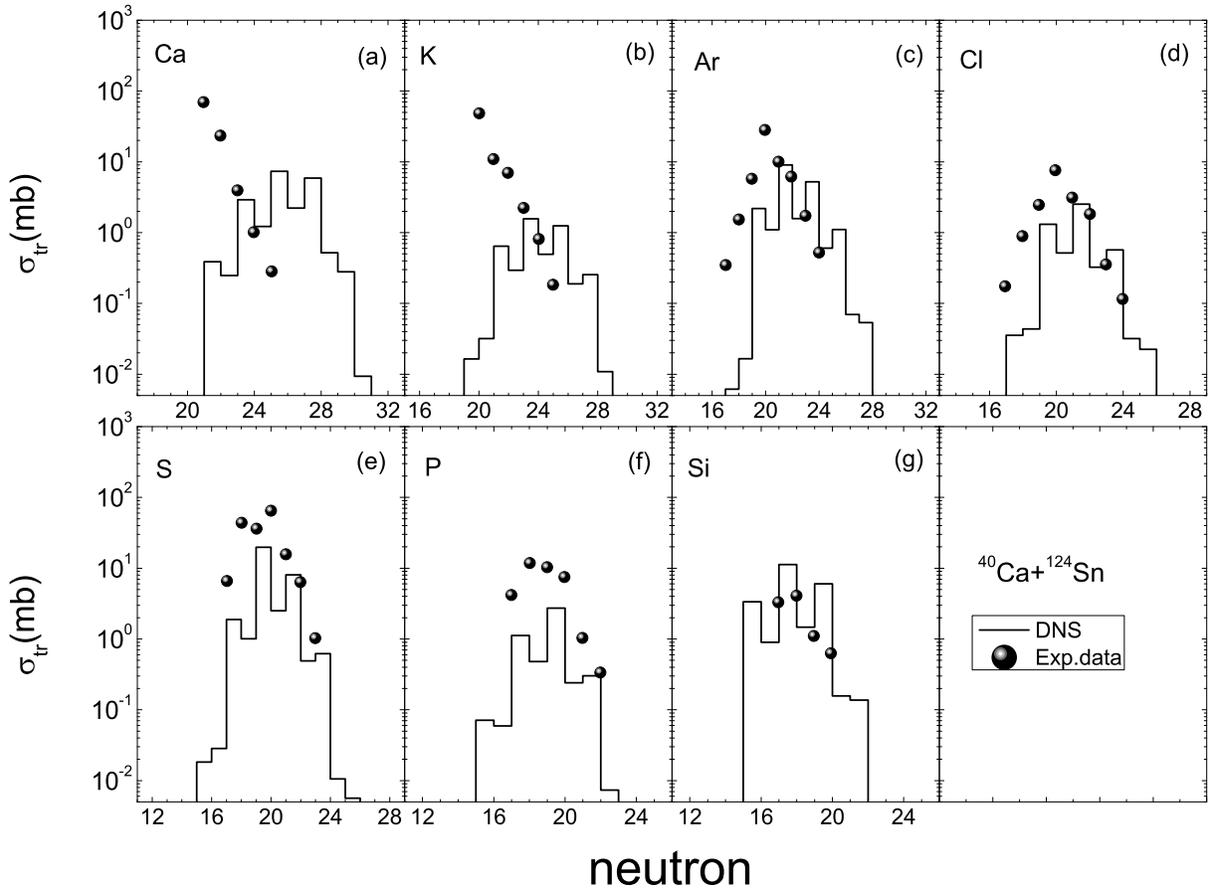}
\caption{\label{fig:wide} Calculated projectile-like fragment (PLF) production as a function of neutron number in the multinucleon transfer reaction of $^{40}$Ca+$^{124}$Sn at the center of mass (c.m.) energy of 128.5 MeV and compared with the available data \cite{Co96}.}
\end{figure*}
%%%%%%%%%%%%%%%%%%%%%%%%%%%%%%%%%%%%%%%%%%%%%%%%%%%%%%%%%%%%%%

%%%%%%%%%%%%%%%%%%%%%%%%%%%%%%%%%%%%% figure 4 %%%%%%%%%%%%%%%%%%%%
\begin{figure*}
\includegraphics[width=16 cm]{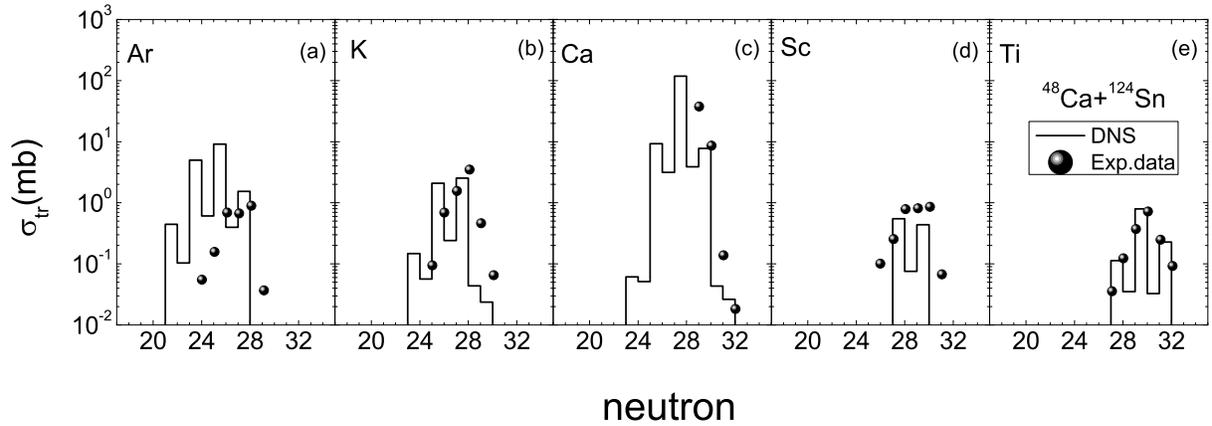}
\caption{\label{fig:wide} The same as in Fig. 3, but for collisions of $^{48}$Ca+$^{124}$Sn at the c.m. energy of 125.4 MeV.}
\end{figure*}
%%%%%%%%%%%%%%%%%%%%%%%%%%%%%%%%%%%%%%%%%%%%%%%%%%%%%%%%%%%%%%

%%%%%%%%%%%%%%%%%%%%%%%%%%%%%%%%%%%%% figure 5 %%%%%%%%%%%%%%%%%%%%
\begin{figure*}
\includegraphics[width=16 cm]{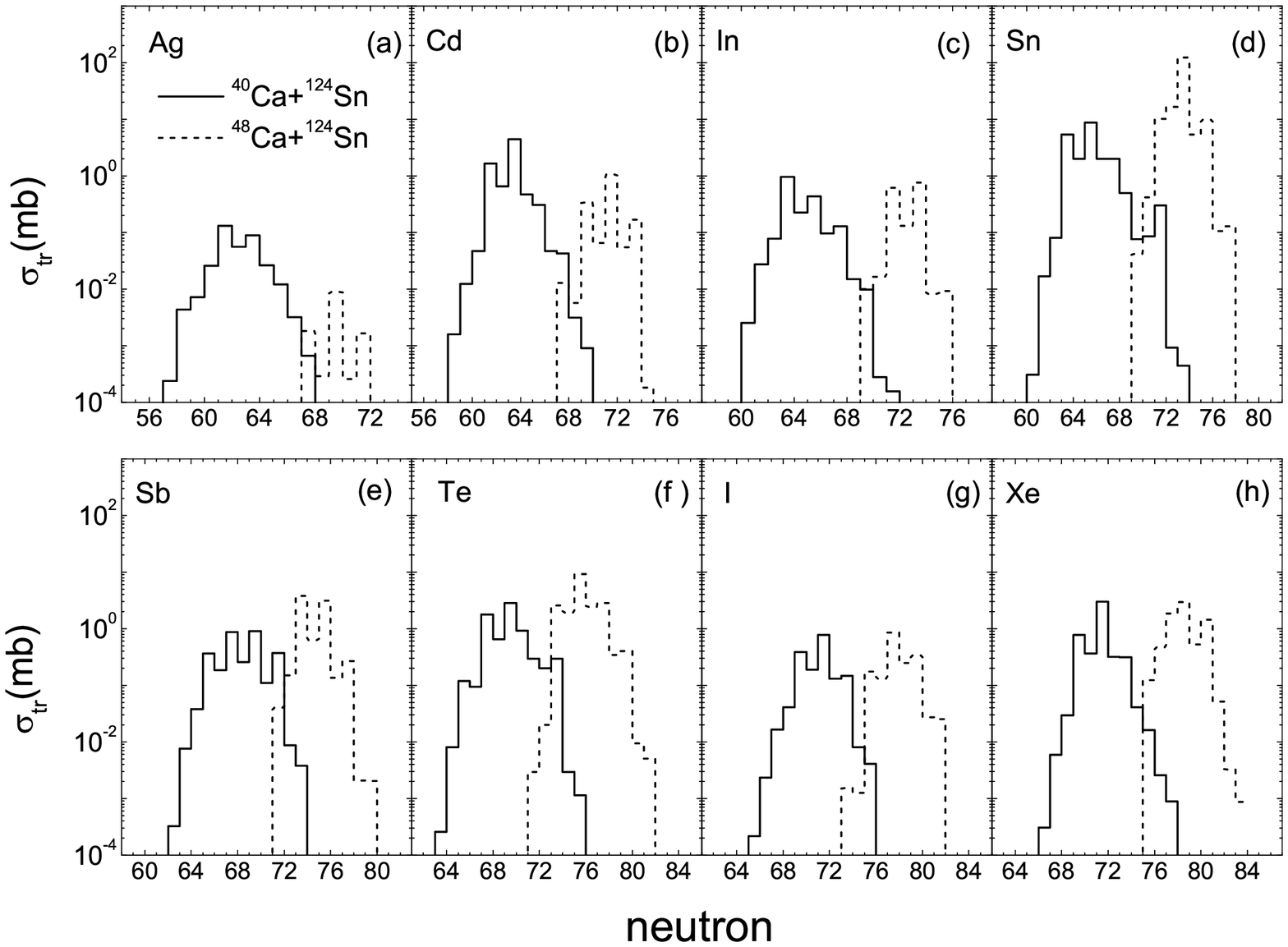}
\caption{\label{fig:wide} Production of the target-like fragments (TLFs) in the $^{40}$Ca+$^{124}$Sn and $^{48}$Ca+$^{124}$Sn reactions at the c.m. incident energies of 128.5 MeV and 125.4 MeV, respectively.}
\end{figure*}
%%%%%%%%%%%%%%%%%%%%%%%%%%%%%%%%%%%%%%%%%%%%%%%%%%%%%%%%%%%%%%

The MNT mechanism has been further investigated with the actinide nuclei based reactions. The system of $^{48}$Ca+$^{248}$Cm was used to produce the neutron-deficient nuclei, i.e., the new isotopes $^{216}$U, $^{219}$Np, $^{223}$Am, $^{229}$Am and $^{233}$Bk \cite{De15}. The investigations are particularly interesting for checking the shell evolution in the domain of proton-rich nuclei besides the new isotope synthesis. Shown in Fig. 6 is the fragment distributions formed in the MNT reactions of $^{40}$Ca, $^{40}$Ar, $^{58}$Ni+$^{232}$Th at the energies of 190 MeV ($V_{C}$=192.2 MeV), 166 MeV ($V_{C}$=166.3 MeV) and 256 MeV ($V_{C}$=255.1 MeV), respectively. The target-like fragments are produced towards the proton-rich side, in particular with the $^{58}$Ni induced reactions. The $^{40}$Ar+$^{232}$Th reaction is favorable for producing the more neutron-rich heavy nuclei because of loosely bound properties for the nuclide $^{40}$Ar. The entrance channel effects in the MNT reactions have been further investigated in the $^{40}$Ca, $^{58}$Ni+$^{238}$U as shown in Fig. 7 and $^{40}$Ca, $^{48}$Ca and $^{58}$Ni bombarding $^{248}$Cm as shown in Fig. 8 around the barrier energies. It should be noticed that the $^{58}$Ni induced reactions have the broader isotope distributions and would be a nice candidate nuclide for creating the new proton-rich actinide nuclei. The neutron-rich projectile nuclei are available for producing the more neutron-rich heavy nuclei. The pick-up reactions are more easily in collisions of $^{40}$Ca+$^{238}$U/$^{248}$Cm, which enable the larger cross sections for the proton-rich nucleus formation via transferred nucleons from target.

%%%%%%%%%%%%%%%%%%%%%%%%%%%%%%%%%%%%% figure 6 %%%%%%%%%%%%%%%%%%%%
\begin{figure*}
\includegraphics[width=16 cm]{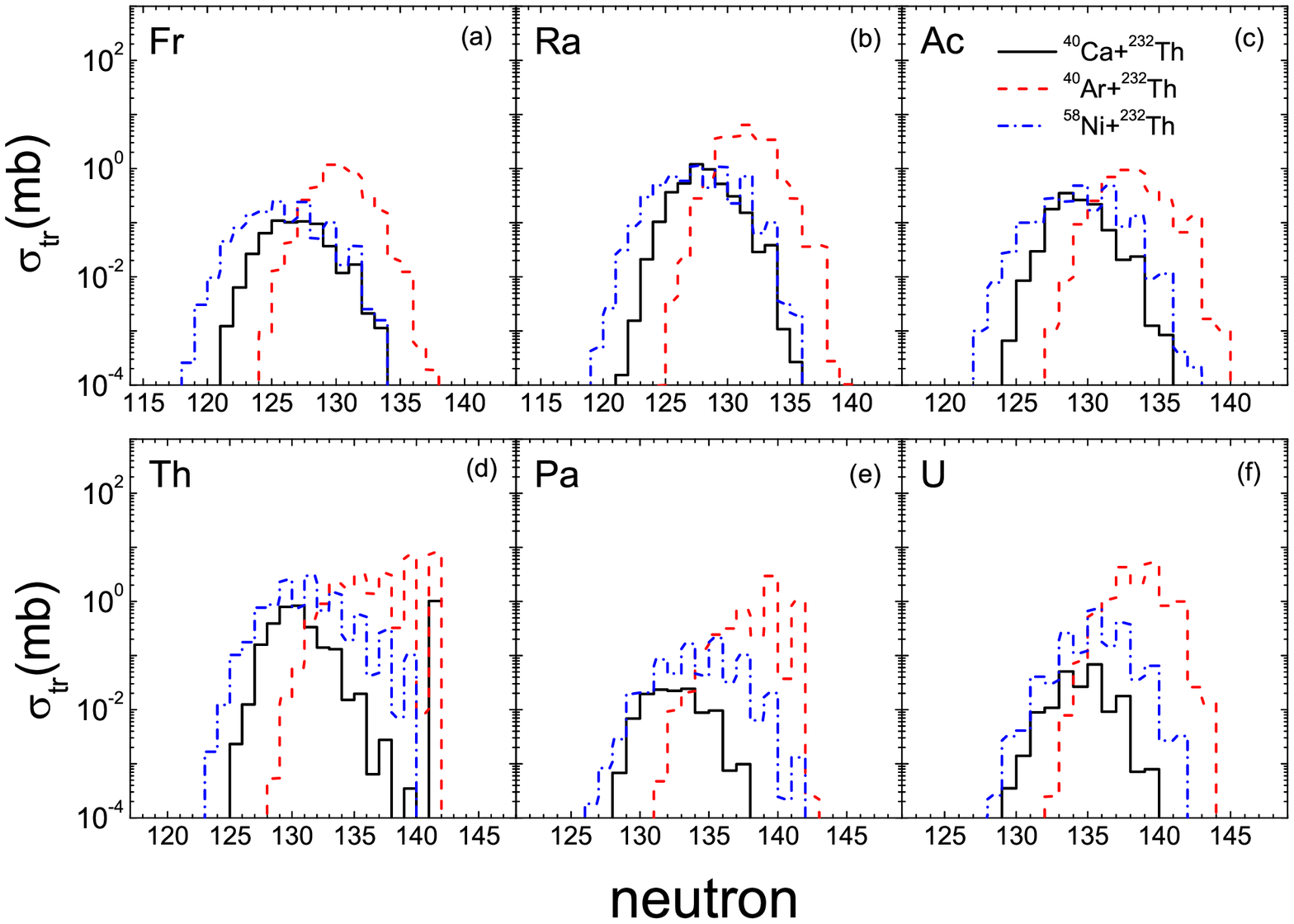}
\caption{\label{fig:wide}(Color online) The fragment production in the collisions of $^{40}$Ca, $^{40}$Ar, $^{58}$Ni+$^{232}$Th at the energies of 190 MeV, 165 MeV and 256 MeV, respectively.}
\end{figure*}
%%%%%%%%%%%%%%%%%%%%%%%%%%%%%%%%%%%%%%%%%%%%%%%%%%%%%%%%%%%%%%

%%%%%%%%%%%%%%%%%%%%%%%%%%%%%%%%%%%%% figure 7 %%%%%%%%%%%%%%%%%%%%
\begin{figure*}
\includegraphics[width=16 cm]{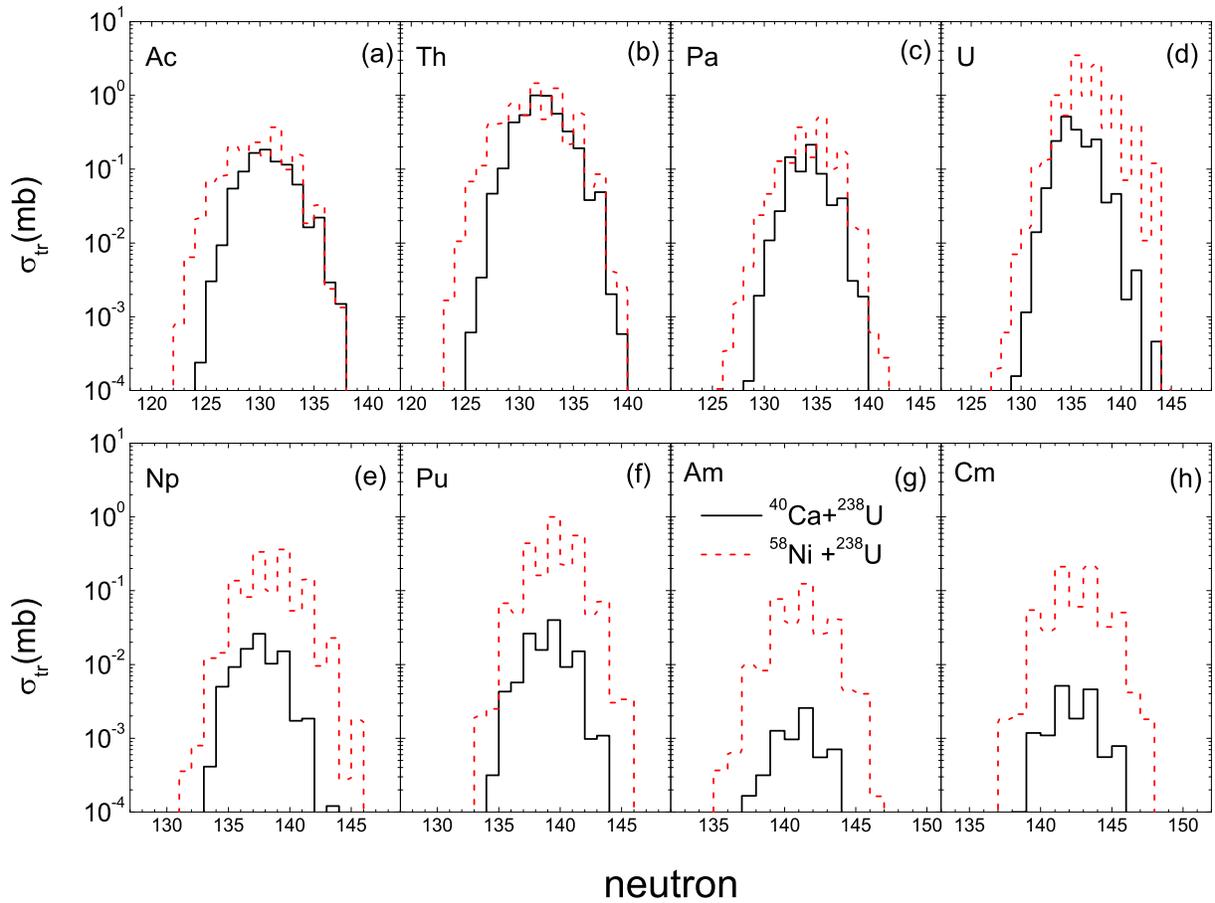}
\caption{\label{fig:wide} The TLF distributions  in the multinucleon transfer reaction of $^{40}$Ca and $^{58}$Ni on the target of $^{238}$U at the c.m. energies of 195.2 MeV and 261.3 MeV, respectively.}
\end{figure*}
%%%%%%%%%%%%%%%%%%%%%%%%%%%%%%%%%%%%%%%%%%%%%%%%%%%%%%%%%%%%%%

%%%%%%%%%%%%%%%%%%%%%%%%%%%%%%%%%%%%% figure 8 %%%%%%%%%%%%%%%%%%%%
\begin{figure*}
\includegraphics[width=16 cm]{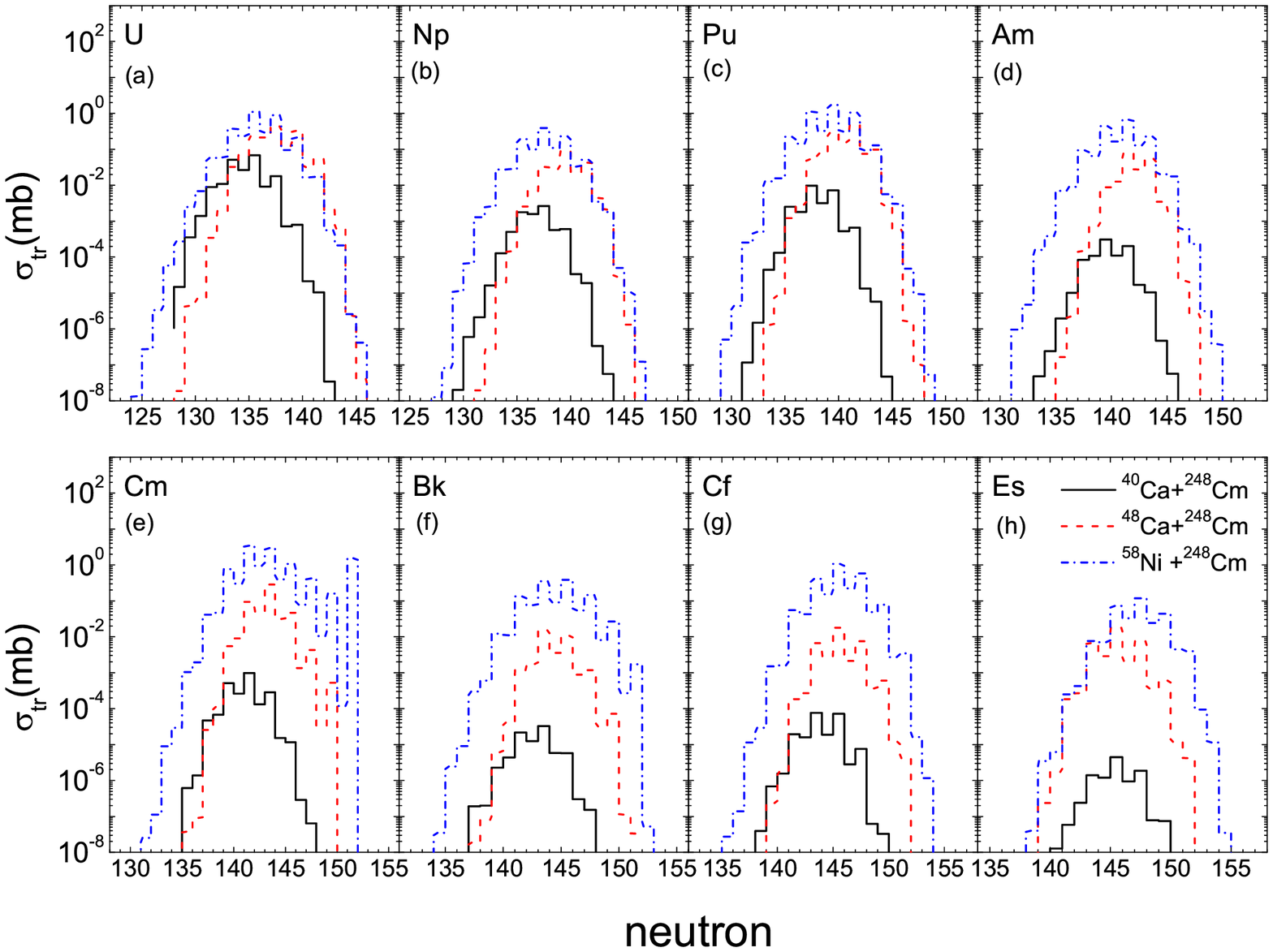}
\caption{\label{fig:wide}(Color online) Distributions of the TLFs in the multinucleon transfer reactions with $^{40}$Ca, $^{48}$Ca and $^{58}$Ni bombarding $^{248}$Cm at the energies of 199.8 MeV, 226.2 MeV and 271.5 MeV, respectively.}
\end{figure*}
%%%%%%%%%%%%%%%%%%%%%%%%%%%%%%%%%%%%%%%%%%%%%%%%%%%%%%%%%%%%%%

The structure of isotope products in the MNT reactions is related to the beam energy. More energy can be dissipated into the DNS with increasing the incident energy of colliding system, which leads to the higher local excitation energy and the larger yields of primary fragments. The excited primary fragments proceed with the deexcitation by evaporating light particles and fission. It was found that the total mass and charge distributions of the survival fragments around the shell closure N=126 weakly depend on the bombarding energy in the MNT reaction of $^{136}$Xe+$^{198}$Pt \cite{Fe17}. The isotopic distributions are further investigated as shown in Fig. 9 in collisions of $^{58}$Ni+$^{248}$Cm. The c.m. energies of 243 MeV, 271.2 MeV and 324.3 MeV are selected to be below, close to and above the Coulomb barrier ($V_{C}=$269.5 MeV), respectively. The production cross sections of isotopes from U to Es are quite different with varying the beam energy. It is interest to be noticed that the energy just around the barrier is optimal in the transfer reactions. The higher beam energy leads to the smaller survival probabilities of primary fragments formed in the MNT reactions. Therefore, the incident energy should be selected close to the Coulomb barrier of colliding system for measuring the fragments in experiments. The neutron shell of N=152 is pronounced in the yields of Am and Cm isotopes. Experiments in the heavy-ion accelerator research facility in Lanzhou (HIRFL) are planning for creating the heavy neutron-rich nuclei via the MNT reactions.

%%%%%%%%%%%%%%%%%%%%%%%%%%%%%%%%%%%%% figure 9 %%%%%%%%%%%%%%%%%%%%
\begin{figure*}
\includegraphics[width=16 cm]{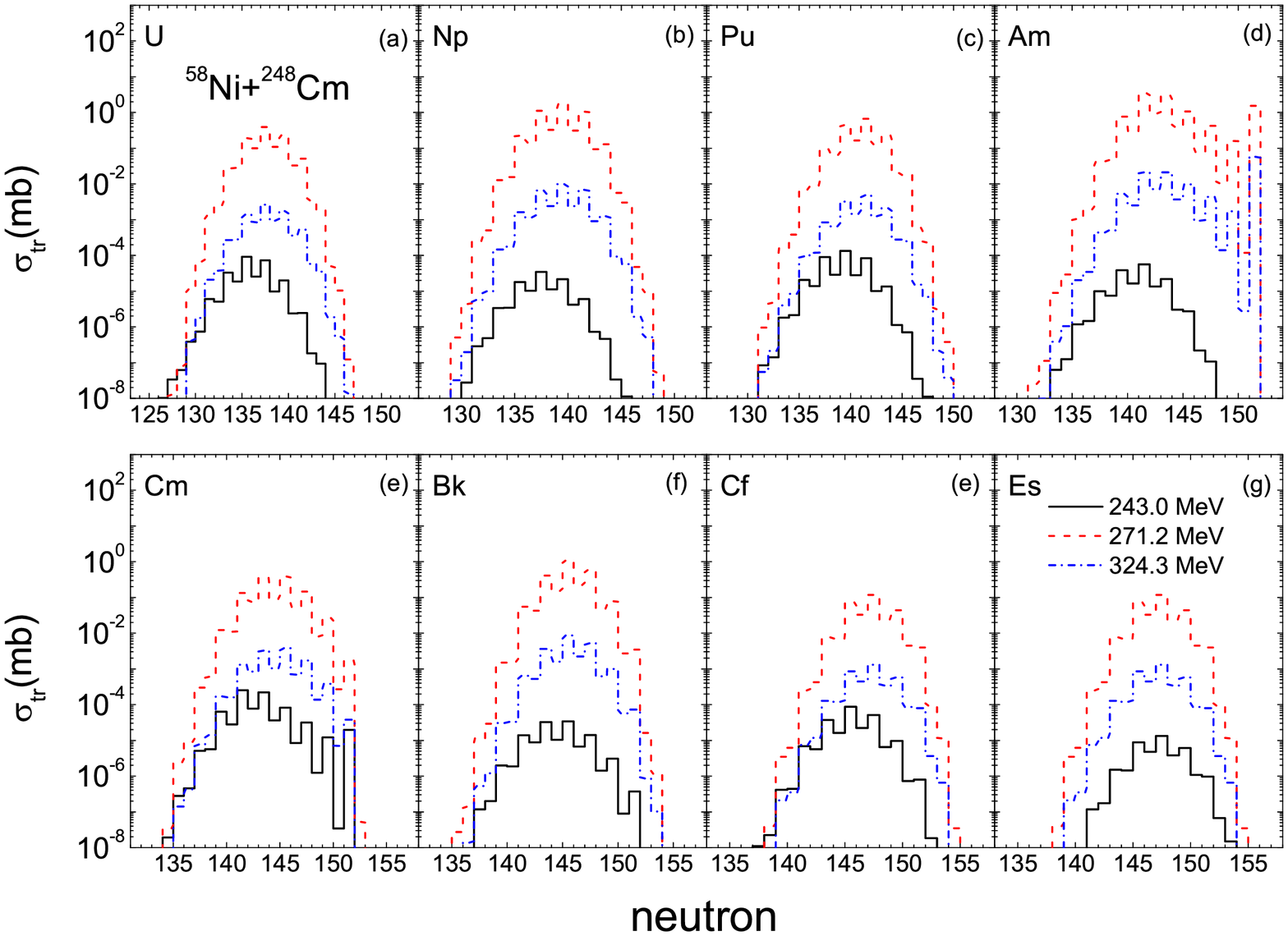}
\caption{\label{fig:wide}(Color online) Incident energy dependence on the production of TLF isotopes in the transfer reactions of $^{58}$Ni+$^{248}$Cm.}
\end{figure*}
%%%%%%%%%%%%%%%%%%%%%%%%%%%%%%%%%%%%%%%%%%%%%%%%%%%%%%%%%%%%%%

\section{Conclusions}

In summary, the multinucleon transfer reactions near Coulomb barrier energies have been investigated within the framework of the DNS model, in which the nucleon transfer takes place after the capture stage of two colliding partners. The all possible orientations in the transfer dynamics are included by the barrier distribution approach. The available experimental data of transferred fragments can be nicely reproduced. The isospin dissipation in the nucleon transfer, relative motion energy and angular momentum relaxation contributes the fragment production away from the $\beta$-stability. The shell structure and odd-even effects play a significant role on the fragment distributions and enhance the heavy neutron-rich isotopes. The neutron-rich projectile nucleus enlarges the isotope distribution and is favorable to produce the heavy neutron-rich isotopes. The projectile-target combinations and the incident energies impact the fragment distributions.

\section{Acknowledgements}

This work was supported by the Major State Basic Research Development Program in China (No. 2014CB845405 and No. 2015CB856903), the National Natural Science Foundation of China (Projects Nos 11675226, 11175218 and U1332207), the Natural and Science Foundation in Henan Province (162300410179) and the Program for the Excellent Youth (154100510007) at Henan Normal University.

\end{document}